\documentclass[traditabstract]{aa} 

\usepackage{graphicx}
\usepackage{txfonts}
\usepackage{natbib}
\bibpunct{(}{)}{;}{a}{}{,} 
\usepackage{xcolor}
\usepackage{amsmath}
\usepackage{nccmath}
\usepackage{comment}
\usepackage{multirow}
\usepackage{soul}
\usepackage{ulem}

\begin{document}

   \title{Discovery of long-period magnetic field oscillations and motions in isolated sunspots}


   \author{A. B. Gri\~n\'on-Mar\'{\i}n\inst{1,2,3}, 
          A. Pastor Yabar\inst{4},
          H. Socas-Navarro\inst{1,2},
          \and
          R. Centeno\inst{5}
          }

   \institute{Instituto de Astrof\'{\i}sica de Canarias,
              V\'{\i}a L\'actea, 38205 La Laguna, Tenerife, Spain \\
         \and
              Universidad de La Laguna, Departamento de Astrof\'{\i}sica, 
              38206 La Laguna, Tenerife, Spain \\
         \and
              W. W. Hansen Experimental Physics Laboratory, 
              Stanford University, Stanford, 
              CA 94305-4085, USA \\
         \and
              Leibniz Institut f\"ur Sonnenphysik (KIS),
              Sch\"oneckstr. 6, 79104 Freiburg, Germany \\
         \and
              High Altitude Observatory (NCAR),
              3080 Center Green Dr. Boulder CO 80301 \\
             }

   \date{Received Month Day, Year; accepted Month Day, 2019}

 
  \abstract
   {
   We analyse the temporal evolution of the inclination component of the magnetic field vector for the penumbral area of 25 isolated sunspots. Compared to previous works, the use of data from the HMI instrument aboard the SDO observatory facilitates the study of very long time series ($\approx$1 week), compared to previous works, with a good spatial and temporal resolution. We used the wavelet technique and  we found some filamentary-shaped events with large wavelet power. Their distribution of periods is broad, ranging from the lower limit for this study of 48 minutes up to 63 hours. An interesting property of these events is that they do not appear homogeneously all around the penumbra but they seem to concentrate at particular locations. The cross-comparison of these wavelet maps with AIA data shows that the regions where these events appear are visually related to the coronal loops that connect the outer penumbra to one or more neighbouring opposite polarity flux patches. 
   }

   \keywords{sunspots – Sun: magnetic fields – Sun: oscillations – Sun: photosphere}
   \titlerunning{short title}
   \authorrunning{name(s) of author(s)}
   \maketitle

%

\section{Introduction}
\label{introduction}
The large-scale magnetic field in the Sun is generated by a large-scale dynamo in the solar interior and driven upwards by plasma buoyancy until it emerges through the surface. The interaction of plasma motions and magnetic fields results in a plethora of magnetohydrodynamical processes, still not fully understood, which give rise to solar activity. In this context, the penumbra of sunspots is a very dynamic environment and many authors have reported different dynamical phenomena (see \citealt{borrero2011}, \citealt{khomenko2015}, and \citealt{tritschler2009} for further information). 

One important dynamical phenomenon is the oscillations that take place in the penumbra. Many authors have studied the presence of these oscillations using different physical parameters such as line-of-sight velocity, continuum intensity, magnetic field strength, inclination, or azimuth. 

\cite{lites1988} studied velocity in sunspots and found that five-minute oscillations are dominant in the outer part of the penumbra. Later, \cite{balthasar1999} confirmed this behaviour. \cite{marco1996} found indications of penumbral oscillations in deep layers of the photosphere with variations between the inner and outer part of the penumbra. The maximum power was located around periods of five minutes. Also, \cite{nagashima2007} reported three-minute oscillations in intensity around the penumbra. These authors also found that the power of photospheric intensity and velocity oscillations was smaller compared to the surrounding quiet Sun. \cite{bellotrubio2000} found that the amplitude of the velocity oscillations increases towards the umbra/penumbra boundary.

\begin{table*}
  \caption{Previous results about long-period oscillations.}
  \label{Tab-previous_results}
  \centering
    \begin{tabular} {l c c} 
      \hline\hline
      Paper & Parameter & Periods \\
       & & (m=minutes, h=hours) \\
      \hline\hline
      \multirow{5}{*}{\citealt{borzov1986}} & \multirow{5}{*}{Magnetograms} & 35 m \\
       &  & 106 m \\
       &  & 124 m \\
       &  & 126 m \\
       &  & 140 m \\
      \hline
      \citealt{nagovitsyna2002} & Photoheliograms & 40 - 100 m \\
      \hline
      \citealt{efremov2007} & $v_{los}$ Maps & 40 - 80 m \\
      \hline
      \multirow{2}{*}{\citealt{solovev2008}} & Magnetograms \& & 40 - 100 m \\
       & $v_{los}$ Maps & 100 - 220 m \\
      \hline
      \citealt{efremov2009} & $v_{los}$ Maps & 60 - 80 m \\
      \hline
      \multirow{2}{*}{\citealt{bakunina2009}} & Microwave images & \multirow{2}{*}{20 - 150 m} \\
       & (5.7 GHz and 17 GHz) & \\
      \hline
      \multirow{5}{*}{\citealt{efremov2010}} &  & 40 - 45 m \\
       & \multirow{2}{*}{Magnetograms \&} & 60 - 80 m \\
       & \multirow{2}{*}{$v_{los}$ Maps} & 135 - 170 m \\
       &  & 220 - 240 m \\
       &  & 460 - 500 m \\
      \hline
      \multirow{2}{*}{\citealt{chorley2010}} & Microwave images & \multirow{2}{*}{16 - 88 m} \\
       & (17 GHz) & \\
      \hline
      \multirow{2}{*}{\citealt{smirnova2011}} & Microwave images & 10 - 60 m \\
       & (37 and 93 GHz) & 80 - 130 m \\
      \hline
      \multirow{6}{*}{\citealt{kallunki2012}} & \multirow{6}{*}{Magnetograms} & 3 - 5 m \\
       &  & 60 - 80 m \\
       &  & 10 - 23 m \\
       &  & 220 - 240 m \\\
       &  & 340 m \\
       &  & 470 m \\
      \hline
      \multirow{6}{*}{\citealt{efremov2012}} & \multirow{6}{*}{Magnetograms} & 40 - 45 m  \\
       &  & 60 - 80 m \\
       &  & 135 - 170 m \\
       &  & 220 - 250 m \\
       &  & 480 - 520 m \\
       &  & 800 - 1300 m \\
      \hline
      \citealt{smirnova2013a} & Magnetograms & 0.5 - 40 h \\
      \hline
      \multirow{2}{*}{\citealt{smirnova2013b}} & Magnetograms \& & \multirow{2}{*}{200 - 400 m} \\
       & (37 GHz) & \\
      \hline
      \multirow{3}{*}{\citealt{abramov2013a}} & Magnetograms \& & 30 - 40 m \\
       & Radio Maps & 70 - 100 m \\
       & (17 GHz) & 150 - 200 m \\
      \hline
      \multirow{4}{*}{\citealt{abramov2013b}} &  & 30 - 40 m \\
       & Magnetograms \& & 60 - 70 m \\
       & Radio Maps (17 GHz) & 100 - 110 m \\
       &  & 150 - 200 m \\
      \hline
      \multirow{2}{*}{\citealt{bakunina2013}} & Microwave images & \multirow{2}{*}{22 - 170 m} \\
       & (5.7 GHz and 17 GHz) & \\
      \hline
      \multirow{3}{*}{\citealt{efremov2014}} & \multirow{3}{*}{Magnetograms} & 10 - 12 h \\
       &  & 32 - 35 h \\
       &  & 35 - 48 h \\
      \hline
    \end{tabular}
    \tablefoot{First column corresponds to the published paper, the second describes the parameter used to obtain the periods of the third column.}
\end{table*}

Several authors have measured fluctuations of the photospheric magnetic field strength in sunspots to find periods of around three to five minutes and amplitudes ranging from a few gauss to tens of gauss. \cite{ruedi1998a} analysed oscillations in the velocity and magnetic field in sunspots. They discovered oscillations located in different parts of the sunspots with an rms of 6.4 G. Other researchers found oscillations concentrated in the umbra or in the penumbra. On the one hand, \cite{balthasar1999} obtained amplitudes of up to 50 G in individual patches of the penumbra. \cite{kupke2000} detected oscillatory behaviour in the longitudinal field strength with an rms of 22 G in the frequency band of five minutes, located in the umbra/penumbra boundary. \cite{zhugzhda2000} detected oscillations in magnetic field strength in small dark patches of the penumbra. \cite{balthasar2003} found small periodic variations in magnetic field strength, inclination, and azimuth located in small areas of the penumbra. By contrast, \cite{landgraf1997} tried to find oscillations in velocity and magnetic field strength within a sunspot umbra, but found that the apparent variations in the magnetic field strength did not exhibit significant oscillations. \cite{lites1998} reported an upper limit of the amplitude of the magnetic field oscillations (4 G) and considered these oscillations to be of instrumental rather than solar origin. However, \cite{bellotrubio2000} studied the magnetic field strength using spectropolarimetric data and obtained fluctuations with an amplitude of about 10 G and a period of five minutes within the umbra. These authors suggested that these oscillations were caused by opacity fluctuations that move the region where the spectral lines are sensitive to magnetic field upwards and downwards.

The above-mentioned oscillations are in the range of a few minutes. Long-period oscillations have also been reported by several authors, who obtained periods between tens of minutes to several hours or days. Such long-period oscillations are difficult to detect because, in addition to good and stable observing conditions, homogeneous and stable instrumentation is also required. Many authors detected long-period oscillations using different parameters such as magnetograms, line-of-sight velocities, microwave, or radio emission maps. Table \ref{Tab-previous_results} summarises the parameters used by multiple authors and the periods of the oscillations reported in these works. These authors found periods between three minutes and 54 hours. Some of these studies suggest that these oscillations can be related to global eigenmodes of the sunspot as a whole. None of the previous studies focussed on oscillations in magnetic field inclination, which is the parameter used in this paper.

In this paper we report on the discovery of a new dynamical phenomenon in the orientation of the photospheric magnetic field vector with characteristic periods and timescales of several hours. In contrast to the long-term oscillations mentioned above, in this case these events appear in particular locations of the sunspot penumbrae. 


\section{Data and methodology}
\label{data}

We used the magnetic data products of 25 alpha sunspots from the Helioseismic and Magnetic Imager (HMI; \citealt{scherrer2012}, \citealt{schou2012}) on board the Solar Dynamics Observatory (SDO; \citealt{pesnell2012}) space mission, which allows a long-term analysis with high cadence (12 minutes), good spatial resolution (0\farcs504), and high time coverage ($\approx$1 week, depending on the analysed sunspot). In particular, we used the Space-weather HMI Active Region Patches (SHARP; \citealt{bobra2014}) data products. In order to get these products, observed Stokes parameters for the \ion{Fe}{I} 6173 \AA\ spectral line are inverted using the Very Fast Inversion of the Stokes Vector code (VFISV; \citealt{borrero2011b}; \citealt{centeno2014}). This code assumes a Milne-Eddington model for the solar atmosphere and solves the radiative transfer equation to derive the magnetic and thermodynamic parameters. After inferring the different atmospheric parameters, the magnetic field vector was disambiguated \citep{barnes2012}, leading to the data products used in this work.

To perform this study we applied the same steps as in the paper \cite{grinonmarin2017} to the data. Briefly, the magnetic field vector was deprojected by transforming from the line-of-sight coordinate reference system to the local solar reference frame applying two reference system rotations. Projection effects were corrected by applying a geometrical transformation (taking into account the latitude and longitude), and finally the time sequence was aligned using the centre of mass of the umbra. After properly applying these steps, we analysed the time variation of the magnetic field inclination at each pixel of the penumbra when the target sunspot is located between -45$^{\circ}$ and 45$^{\circ}$ of longitude.

\begin{figure}
  \sidecaption
    \includegraphics[angle=90,width=0.5\textwidth]{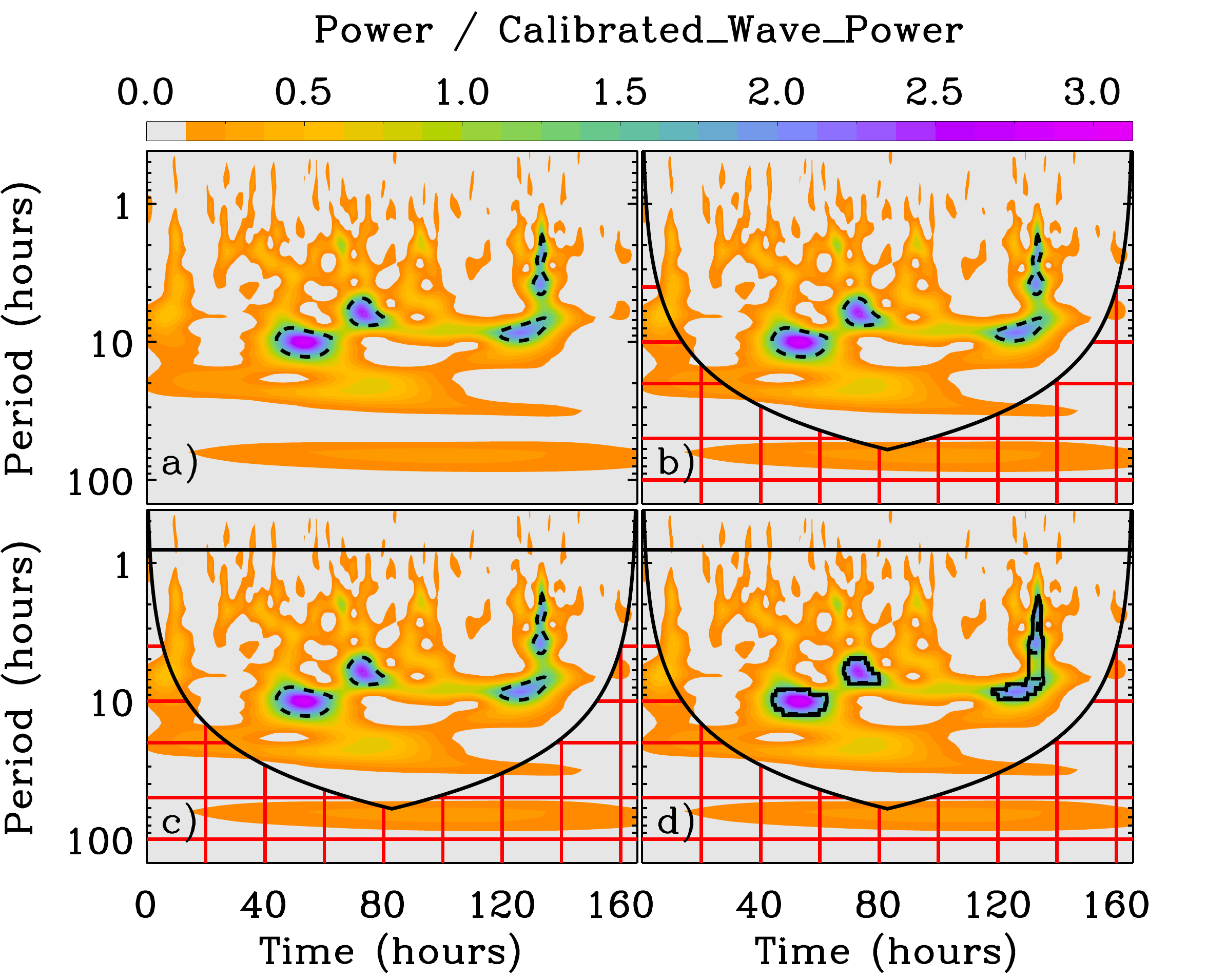}
    \caption{Wavelet decomposition of the time series for one pixel. The background is the same for the four panels and the colour scale is the wavelet power normalised by the calibrated wavelet power (five times the estimated noise level of the magnetic field inclination). The $x$-axis corresponds to the time sequence and the $y$-axis is the period. The horizontal line, dashed line, solid contours, and  red gridded area reflect the different criteria for the event selection explained in Sect. \ref{data}.}
    \label{Fig:wavelet}
\end{figure}

To search for possible motions or oscillations in the inclination of the magnetic field lines, we applied a wavelet analysis using the Interactive Data Language program provided by \cite{torrence1998}. Wavelet analysis is a tool for analysing localised and transient signals within a time series. With this tool, we can retrieve the power spectrum for a time series as a function of time and frequency. This analysis was applied only over the penumbral pixels as the inclination in this region is more reliable than that of the surrounding quiet Sun area. Thus, we can apply the wavelet analysis with confidence that wavelet power is due to actual variations of the magnetic inclination and not associated with noise. Also, the magnetic field inclination of the umbral pixels hardly changes with time and does not present as strong perturbations as the penumbral pixels show. Therefore, we focussed this study on the penumbral pixels. Finally, to select the most significant events and to make their characterisation as accurate as possible, we only analysed the events that simultaneously fulfil all of the following restrictive conditions. The first four criteria are imposed for each pixel of the penumbra individually (illustrated in the four panels of Fig. \ref{Fig:wavelet}) and the fifth is imposed for the events that fulfil the previous four.

\begin{enumerate}[1)]
    \item The periods and times for which the observed wavelet power exceeded the power owing to a five-degree oscillation in inclination were selected. This value corresponds to five times the uncertainty of the magnetic field inclination value in sunspot penumbrae when the sunspot crosses the central meridian ($\approx4\ \sigma$ when the sunspot is at the eastern limb). We chose this threshold to be sure that the detections correspond to significant signals. We made use of the line-of-sight inclination error maps provided by the HMI team to determine the confidence level \citep{bobra2014}. The dashed contours in panel a) of Fig. \ref{Fig:wavelet} indicate the selected areas.

    \item Periods below the wavelet confidence limit were discarded. This cone of influence is the region of the wavelet spectrum in which edge effects become important. Panel b) in Fig. \ref{Fig:wavelet} shows this discarded area, which indicated with a red grid. This cone of influence indicates the maximum period of useful information at each particular time and period. Below this limit the periods are subject to edge effects.

    \item To avoid spurious signals, we followed a restrictive sampling threshold discarding the periods below 48 minutes, which corresponds to two complete periods of a signal oscillating at the Nyquist frequency (the time cadence of the used data is 12 minutes). This condition is shown in panel c) of Fig. \ref{Fig:wavelet} by a horizontal line. This way periods above the horizontal line are discarded.

    \item All occurrences (individual spatial locations and times) in the data that exhibited excess power in one or more frequency ranges were identified. This condition is shown by solid contours in panel d) of Fig. \ref{Fig:wavelet}.
    
    \item A last criterion is added to include some spatio-temporal coherence: a valid event must be observed simultaneously in (at least) three adjacent pixels and have a duration of four or more consecutive frames.
\end{enumerate}

\begin{figure*}
  \sidecaption
    \includegraphics[width=0.7\textwidth]{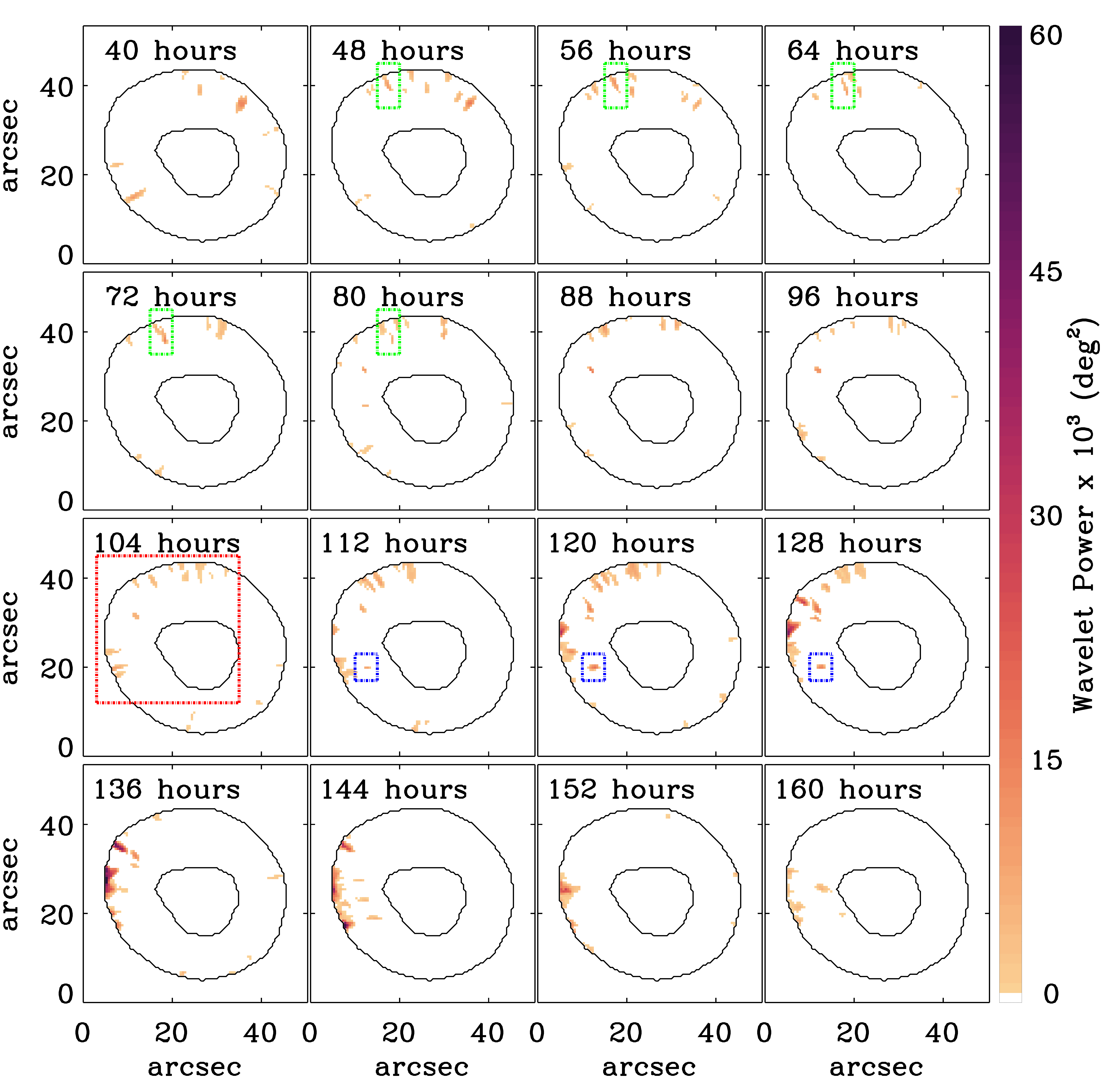}
    \caption{Catalogue of events of the same isolated sunspot in NOAA AR12218 that fulfil the conditions applied for each penumbral pixel. Each panel represents a time step (the time from the beginning of the sequence, in hours, is written in each panel) of the full time sequence. The contours highlight the boundary of the penumbra of the sunspot and the colour scale indicates the wavelet power values of each pixel that meets the imposed selection criteria. The first and last time step are denoted in Fig. \ref{Fig:wavelet_tiempo} with vertical dashed purple lines. The coloured boxes highlight areas of interest as mentioned in the text.}
    \label{Fig:catalogo_eventos}
\end{figure*}

In this manner, we built a catalogue of events that is the basis of our study, which is analysed in the next section.


\section{Analysis and results}
\label{results}

The results presented in the following refer to the sunspot of NOAA AR12218, which has been taken as a representative of the sample described in \cite{grinonmarin2017}. The results of the other sunspots of the sample are shown in Appendix \ref{fenoDinaRes}. The events gathered in the above-mentioned catalogue have a transient character, as they appear and disappear along the whole time sequence considered. Figure \ref{Fig:catalogo_eventos} shows the wavelet power found for the various events present at 16 different timesteps of the time sequence. These events are elongated (see for instance the events indicated by a red rectangle on the 104-hour time step of Fig. \ref{Fig:catalogo_eventos}) and closely aligned with the radial direction from the sunspot centre. The spatial resolution of HMI does not allow us to resolve the penumbral filaments clearly. These filamentary structures do not appear to be related to the photospheric penumbral filaments, as these filamentary structures are wider and do not span the entire penumbra in the radial direction. Also, they are located along filamentary structures localised predominantly near to the outer edge of the penumbra and we find that these phenomena are very dynamic. Some events appear and disappear in $\approx$16 hours (events indicated by a blue rectangle on the 112-128-hour time steps of Fig. \ref{Fig:catalogo_eventos}) and others are more durable living more than 32 hours (events highlighted by a green rectangle on the 48-80-hour time steps of Fig. \ref{Fig:catalogo_eventos}). 

\begin{figure*}
  \begin{center}
    \includegraphics[width=1\textwidth]{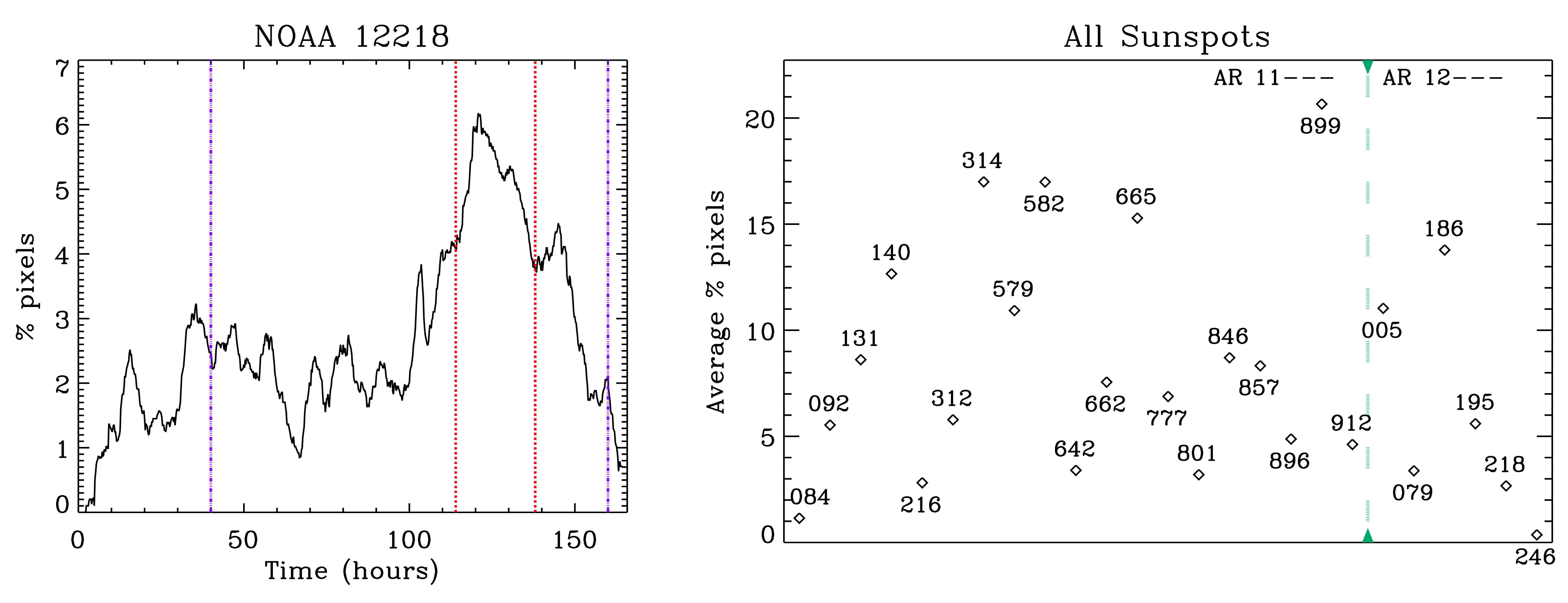}
    \caption{Fraction of penumbral area that satisfies the imposed conditions (see Sect. \ref{data}). The left panel represents this fraction for the sunspot of NOAA AR12218 for each time step. The vertical dashed purple lines indicate the first and last panels of Fig. \ref{Fig:catalogo_eventos} and the red lines indicate the period of time, used as an example in the text, when the percentage of penumbral fraction suffers a large variation. The right panel shows the averaged penumbral fraction for each sunspot of the sample. The vertical dashed green line denotes the separation between the NOAA active regions whose number starts with 11 and those that start with 12.}
    \label{Fig:wavelet_tiempo}
  \end{center}
\end{figure*}

These events cover a varying fraction of the penumbra as individual patches become excited at a given moment, stayed excited for a few hours, and then decreased in power until they disappeared. The left panel of Fig. \ref{Fig:wavelet_tiempo} depicts the penumbral fractional area covered (in percentage) by these events for the sunspot of NOAA AR12218. This percentage varies between 0\% and 6\% and there are large and sudden variations in the area covered by these events (see for instance between t=114 hours and t=138 hours, indicated with two vertical dashed red lines) emphasising their transient character. Also, the presence of these events at the very beginning and end of the time series (first five and last five hours) is smaller. This is likely because during these intervals, the wavelet analysis is valid in a very small frequency range (see panel b in Fig. \ref{Fig:wavelet}) and so, the number of detections is expected to be smaller. The averaged fractional area covered by each sunspot of the sample varies between $\approx0\%$ (NOAA AR12246) to $\approx20\%$ (NOAA AR11899) (see right panel in Fig. \ref{Fig:wavelet_tiempo}). This means that on average, at most $20\%$ of the penumbral area suffers the dynamical phenomena reported in this paper, i.e. these perturbations are not related to global perturbations of the sunspot as a whole, as the reported by other authors (see Sect. \ref{introduction} for references).

\begin{figure}
  \begin{center}
    \includegraphics[width=0.5\textwidth]{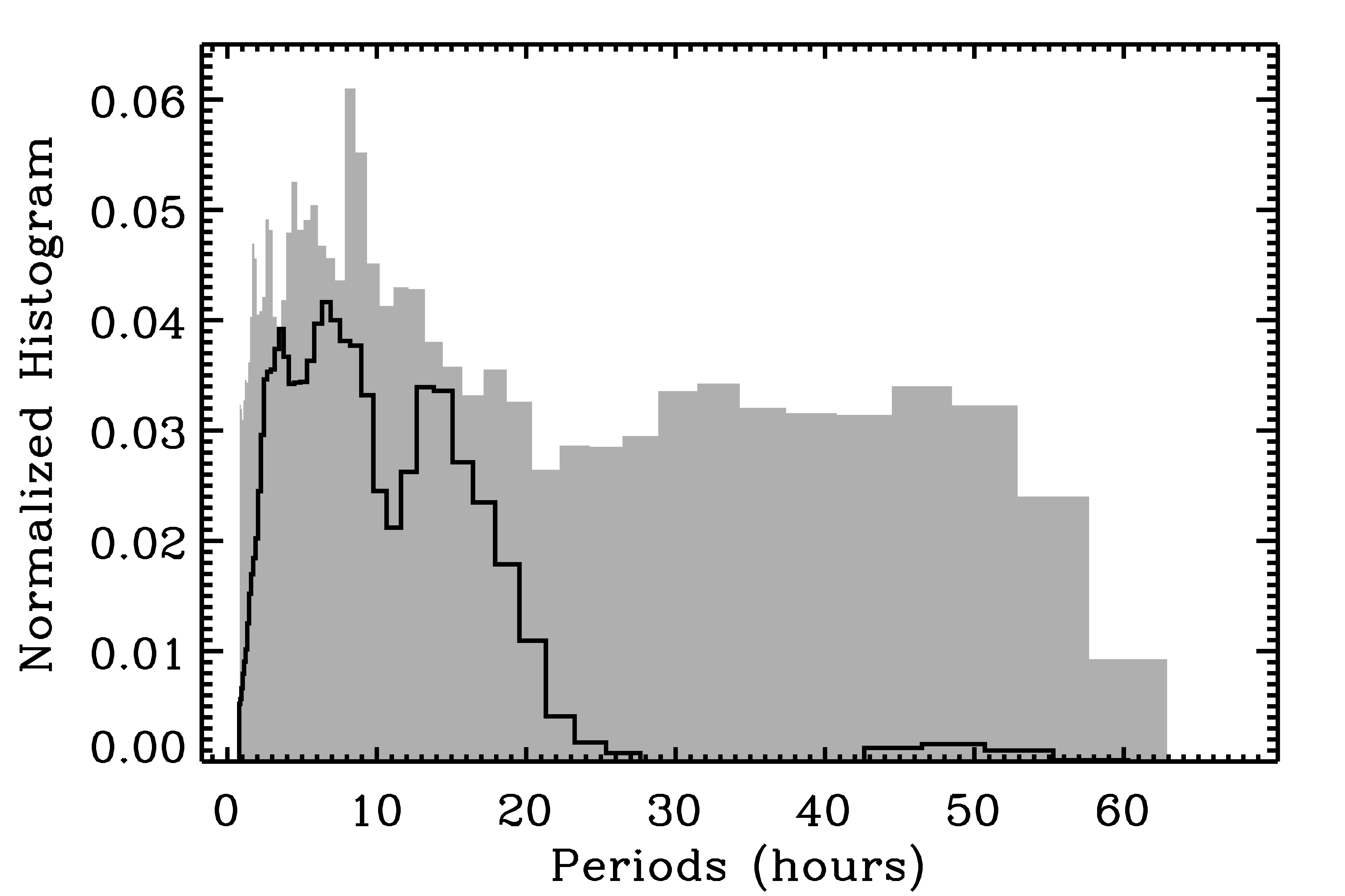}
    \caption{Period distribution of the events identified for NOAA AR12218. The y-axis is the number of pixels that fulfil the applied conditions for different periods. The grey area represents the distribution of the minimum and maximum periods for the complete sample of sunspots.}
    \label{Fig:frecuenciasExcitadas}
  \end{center}
\end{figure}

\begin{figure*}
  \begin{center}
    \includegraphics[width=1\textwidth]{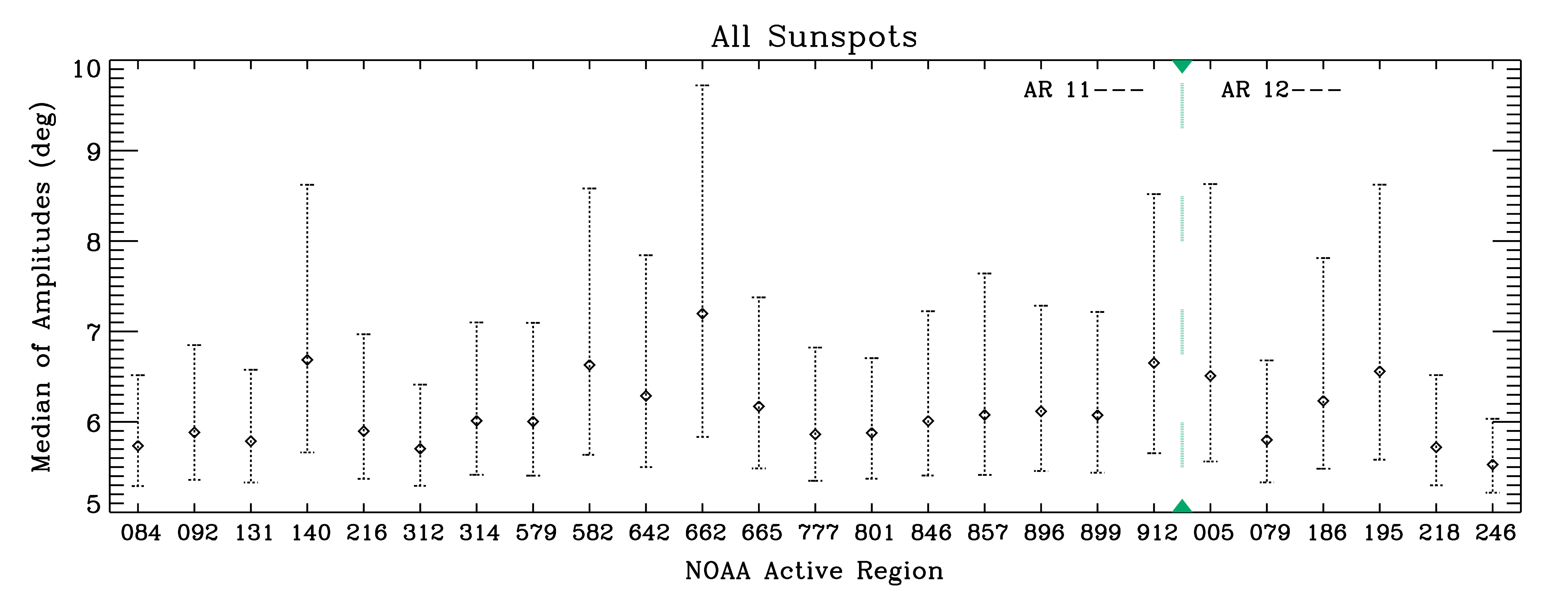}
    \caption{Median of amplitude values detected for each sunspot of the sample. The vertical dashed black lines represent the 25 and 75 percentiles. The vertical dashed green line indicates the separation between the NOAA active regions whose number starts with 11 and those that start with 12.}
    \label{Fig:amplitudes}
  \end{center}
\end{figure*}

Analysing the period distribution of the events of Fig. \ref{Fig:frecuenciasExcitadas} (black solid line), it seems that the analysed sunspot (NOAA AR12218) predominantly exhibits events with wavelet powers in high frequencies; these events have periods of several hours, always higher than 48 minutes, which is one of the imposed conditions. The shaded grey region shows the distribution of minimum and maximum periods for the complete sample of sunspots. There is no clear period value for these events. For example, the distribution of the analysed sunspot (solid dark line) is narrow and seems to be concentrated in the shorter periods. Yet, if we consider the whole sample of sunspots (shaded grey area), it shows a broad distribution with more or less the same probability, i.e. there is neither a characteristic period nor is there any absent frequency. Regarding the oscillatory amplitudes detected we obtained values between $5^{\circ}$ (the minimum value imposed) and $10^{\circ}$. Figure \ref{Fig:amplitudes} shows the median of amplitudes for each sunspot together with 25 and 75 percentiles indicated with vertical bars.

\begin{figure*}
    \sidecaption
    \includegraphics[width=0.7\textwidth]{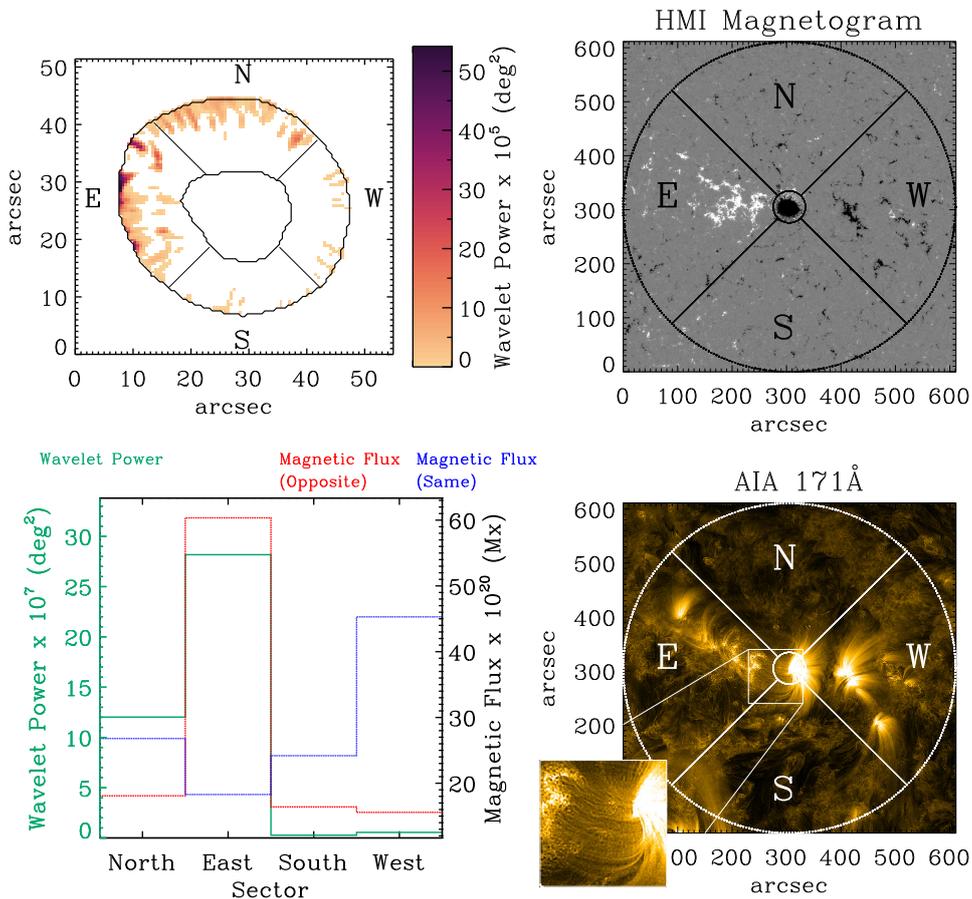}
    \caption{Penumbral distribution of the wavelet power of NOAA AR12218 and its comparison to the spatial distribution of the magnetic field and coronal emission. The upper left panel represents the spatial distribution of events, where the penumbra is divided into four $90^{\circ}$ sectors (north, east, south, and west). The upper and lower right panels show the HMI \mbox{line-of-sight} magnetogram (saturated between \mbox{-200} and 200 gauss) and the corresponding coronal image in the AIA 171~$\AA$ filter (with the area denoted with a white square zoomed in the lower left corner of the image). Both display an extended field of view at the time the sunspot was crossing the central meridian. Finally, the lower left panel represents a histogram of the integrated azimuthal distribution of magnetic fluxes and wavelet power. The black and red lines show the magnetic flux around the target sunspot of the same and opposite polarity, respectively, and the blue line corresponds to the wavelet power of the studied sunspot.}
    \label{Fig:results12218}
\end{figure*}

We then calculated the spatial distribution of events weighted by their integrated power in the sunspot to obtain the results shown in Fig. \ref{Fig:results12218} (upper left panel). We divided the penumbral area into four $90^{\circ}$ sectors, labelled N (north), E (east), S (south), and W (west). In this particular case, most of the power was concentrated in the east sector (and some areas in the north sector), which is the one that faces the opposite polarity of the active region. We compared the integrated event power with the distribution of the magnetic flux in the surroundings of the target sunspot for the aforementioned sectors to quantify this relation. We selected an annular area of the line-of-sight magnetogram around the sunspot to calculate the distribution of the magnetic
flux in the surroundings of the sunspot. The inner radius chosen was such that we excluded the target sunspot, and the radius of the outer circle was such that the full active region was included (see upper right panel of Fig. \ref{Fig:results12218}). Then, we divided this annular area into four $90^{\circ}$ sectors and calculated the magnetic flux taking into account the pixel area projected onto the solar surface. Finally, we computed the total positive and negative magnetic flux contribution for each sector separately. The lower left panel of Fig. \ref{Fig:results12218} shows this comparison. A strong correspondence is found between the location of the largest integrated wavelet power in the penumbra and the general distribution of the closest magnetic flux concentrations of opposite polarity to the target sunspot.

The connection of the observed dynamical phenomena with opposite polarity magnetic concentrations may be studied in chromospheric and coronal images because the large-scale magnetic fields connecting opposite polarity patches in active regions are able to reach higher atmospheric layers and light up in coronal emission lines. To incorporate the additional coronal information we compared the wavelet power sector distribution to coronal images from the Atmospheric Imaging Assembly (AIA; \citealt{lemen2012}) 171~$\AA$ filter (lower right panel of Fig. \ref{Fig:results12218}). These images suggest that those areas with higher power in the photospheric magnetic field inclination are connected to opposite polarity magnetic patches by coronal loops; see the east sector of the target sunspot in the lower right panel of Fig. \ref{Fig:results12218} (denoted by a white square), from which most of the coronal loops emanate and connect with the opposite polarity of the active region.

We repeated the same procedure for 24 other sunspots (see Fig. A.2 and Table A.1 of \citealt{grinonmarin2017}), spanning five years of the solar cycle 24. Examination of the other spots (see the results in Appendix \ref{fenoDinaRes}) in our sample shows a strong correspondence between the location of the events in the penumbra and the general direction to the closest magnetic flux concentrations of opposite polarity. We found that 64\%\footnote{Active regions NOAA 11084, 11092, 11131, 11140, 11314, 11579, 11642, 11662, 11801, 11846, 11857, 11899, 11912, 12005, 12195 and 12218} of the sunspots analysed in this study present the highest power in the sector where most of the opposite polarity magnetic flux is found. In 24\%\footnote{Active regions NOAA 11216, 11665, 11777, 11896, 12079 and 12246} of the sunspots analysed in this work, high oscillatory power is found in the adjacent sector where the opposite polarity magnetic flux is maximum. Also, in 8\%\footnote{Active regions NOAA 11312 and 11582} of the cases the peak of the wavelet power matches the sector with an amount of magnetic flux higher than 80\% of the magnetic flux peak. Only 4\%\footnote{Active regions NOAA 12186} of the sunspots analysed in this work show no correspondence between power and opposite polarity flux. This strong association suggests that the events we observed are related to the motions of magnetic field lines connecting the outer penumbra to one or more neighbouring opposite polarity magnetic flux patches.

\begin{figure}
  \begin{center}
    \includegraphics[width=0.5\textwidth]{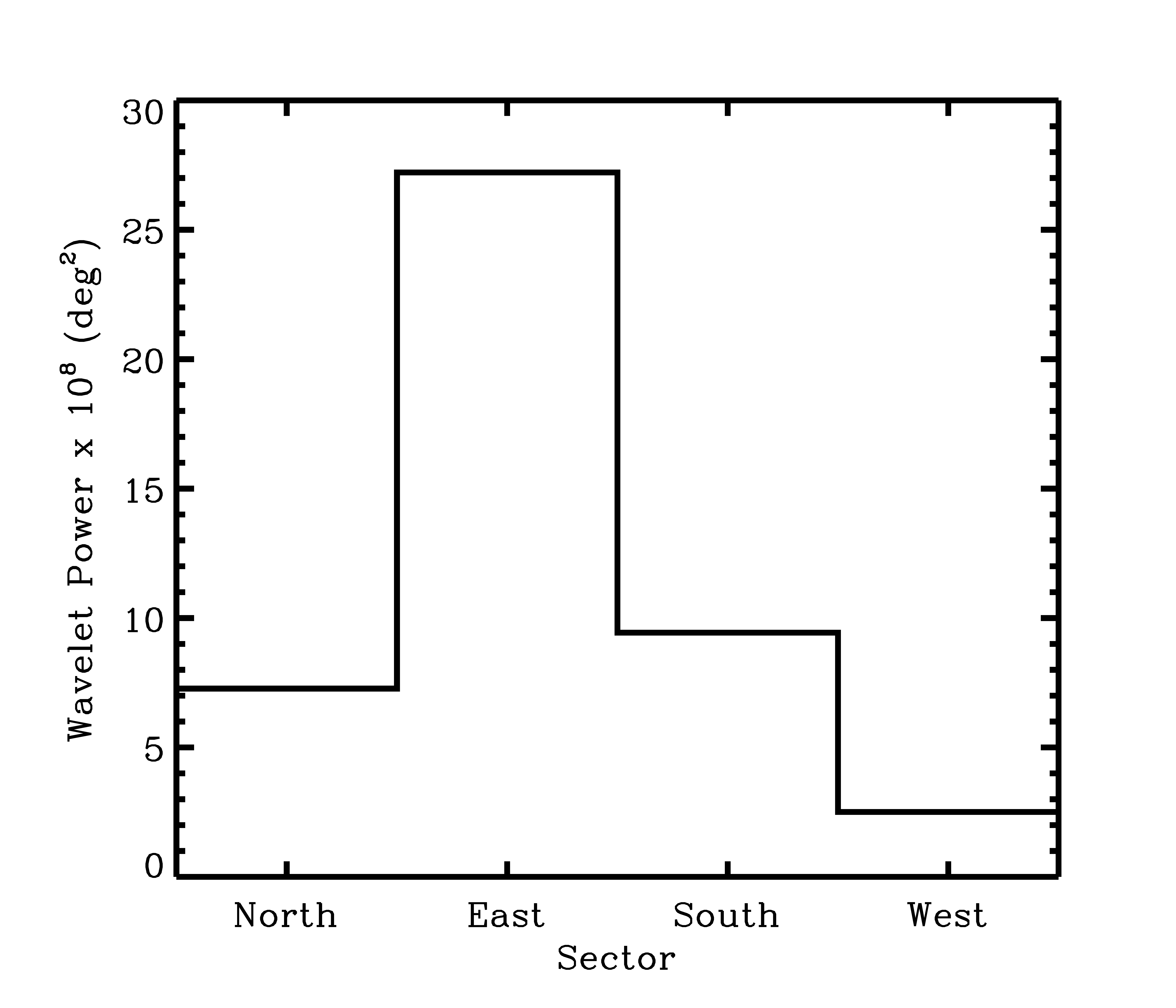}
    \caption{Histogram of the wavelet power average of the 25 sunspots of the studied sample for each selected sector.}
    \label{Fig:histograma}
  \end{center}
\end{figure}

Also, the sample that we analysed for this work considers isolated active regions only, which implies that most of the event detections are located on the east side of the spot, where the following polarity of the active region is found. This is seen clearly in Fig. \ref{Fig:histograma}, which represents the integrated wavelet power of each sector averaged over the entire sample of 25 sunspots analysed. Most of the wavelet power is in the east sector, which is dominated by the opposite polarity flux.

A similar comparison applied over the AIA images suggests that those areas with higher power in the photospheric magnetic field inclination are often connected to opposite polarity flux patches by coronal loops, either in the same active region, as in the lower right panel of Fig. \ref{Fig:results12218}, or in a neighbouring lower right polarity patch as in NOAA AR12246 (see Fig. \ref{Fig:results12246}). Visual comparison of the images of the Appendix \ref{fenoDinaRes} shows that in the 60\%\footnote{Active regions NOAA 11084, 11131, 11314, 11582, 11642, 11662, 11801, 11846, 11896, 11899, 11912, 12005, 12195, 12218, 12246} of the analysed sample the sunspot is apparently connected by coronal loops with the sector where most of the opposite polarity magnetic flux is found. In 24\%\footnote{Active regions NOAA 11092, 11312, 11665, 11777, 11857} of the cases the evidence is inconclusive. Finally, there is no apparent connection by coronal loops in 16\%\footnote{Active regions NOAA 11140, 11579, 12079, 12186} of the sample.

\begin{figure*}
  \sidecaption
    \includegraphics[width=0.7\textwidth]{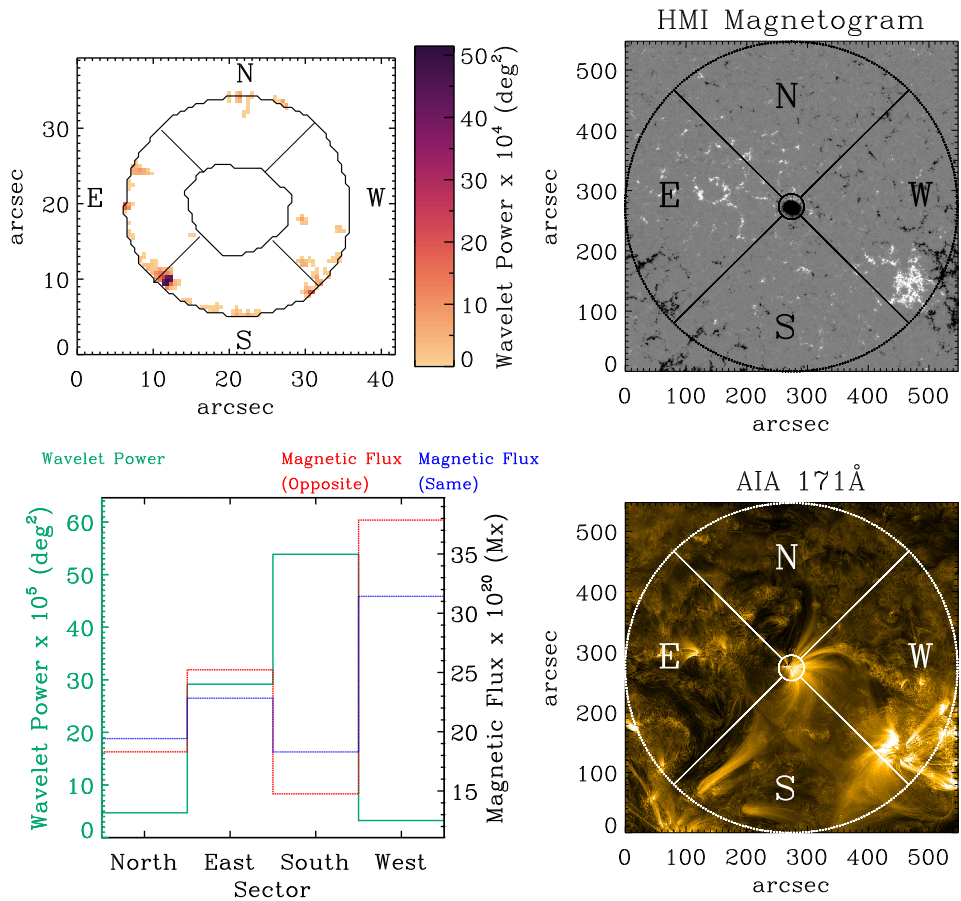}
    \caption{Same as Fig. \ref{Fig:results12218} for the NOAA AR12246.}    
  \label{Fig:results12246}
\end{figure*}

A caveat to the above discussion is that we are comparing dynamic events that occurred over the span of several days with a single magnetogram and one AIA image taken when the sunspot was crossing the central meridian. This is a reasonable approach because, by sample selection, we are studying very stable sunspots throughout their passage across the solar disc. Hence, a single magnetogram should be representative of the whole time series. However, in the upper layers, this might not be the case since the dynamism of the corona might involve strong changes even for the selected sunspots. In any case, we still find a tight connection between event location in the spot and the direction of the opposite polarity and coronal loops.

In some cases, the events detected in our data have well-defined periods, whereas in others we observe a broad range of oscillatory components. We cautiously suggest that this could be caused by the presence of multiple field strands with varying length within the modest HMI resolution element. If each of these strands were oscillating with its own characteristic frequency, we would end up observing a superposition of them. It is also worth emphasising that the motions last only for a few periods, which suggests the presence of an efficient damping mechanism.


\section{Conclusions}
\label{conclusions}

This study led to the discovery of long-period (hours) small-scale perturbations in the magnetic field that occur predominantly in the areas of the penumbra facing the opposite magnetic polarity. Previous works have studied the dynamism of sunspots (see \citealt{borrero2011}, \citealt{khomenko2015}, and \citealt{tritschler2009}), analysing the oscillations that take place in the penumbra. These studies measured oscillations in sunspots as a whole, rather than oscillations localised in small regions of the penumbra, such as those we report on for the first time. The perturbations detected in this paper are characterised by long periods and small spatial scales. This result has been obtained studying the long-term evolution of the local inclination angle of the magnetic field lines of the penumbrae of 25 isolated sunspots. The detected perturbations do not occur homogeneously over the entire sunspot, but are localised along filamentary structures concentrated in specific areas of the penumbra. Moreover, images from the corona suggest a possible association with ultraviolet emission (coronal loops) in the upper atmosphere at the same locations. Because of the very long timescales, it has not been possible to observe these motions before SDO, which brings unprecedented high-cadence, long-term, high-resolution magnetic field observations.

The finding of new dynamical phenomena of an oscillatory nature, such as that reported in this paper, has often resulted in valuable novel diagnostics for the solar plasma because it seems that the photosphere and the solar corona are connected to produce these motions. However, where does the energy of these movements come from? Our speculation is that the magnetic field lines suffer reconnection in the upper layers of the solar atmosphere, and that this phenomenon propagates waves towards the footpoints of the magnetic field lines. In this way, the effects produced by these waves can be detected in the photosphere, in particular, in the penumbral filaments of the sunspots. 



\begin{acknowledgements}
The authors are grateful to the SDO/HMI team for their data. ABGM acknowledges Fundaci\'on La Caixa for the financial support received in the form of a Ph.D. contract. This work was supported by NASA Contract NAS5-02139 (HMI) to Stanford University. The National Center for Atmospheric Research (NCAR) is sponsored by the National Science Foundation. The authors gratefully acknowledge support from the Spanish Ministry of Economy and Competitivity through project AYA2014-60476-P (Solar Magnetometry in the Era of Large Solar Telescopes). Financial support by the German Government through DFG project ``STOK3D Three dimensional Stokes Inversion with magneto-hydrostationary constraints'' is gratefully acknowledged. Wavelet software was provided by C. Torrence and G. Compo, and is available at URL: http://paos.colorado.edu/research/wavelets/.
\end{acknowledgements}


\bibliographystyle{aa}
\bibliography{articulos} 


\begin{appendix} 

\section{Long-Term motions in sunspots: Results for the 25 sunspot database}
\label{fenoDinaRes}
In this appendix we present the results of the rest of the sunspot sample used in this paper. Each figure corresponds to the penumbral distribution of the wavelet power and its comparison to the spatial distribution of the magnetic field and coronal emission. The upper left panel represents the spatial distribution of events, where the penumbra is divided into four $90^{\circ}$ sectors (north, east, south, and west). The upper and lower right panels show the HMI line-of-sight magnetogram and the corresponding coronal image in the AIA 171~$\AA$ filter. Both display an extended FOV at the time the sunspot was crossing the central meridian. Finally, the lower left panel presents a histogram of the integrated azimuthal distribution of magnetic flux and wavelet power. The blue and red lines show the magnetic flux around the target sunspot of the same and opposite polarity, respectively, and the green line corresponds to the wavelet power of the sunspot studied. These plots were calculated following the procedures explained in Sect. \ref{results}.

\begin{figure}
  \begin{center}
    \includegraphics[width=0.5\textwidth]{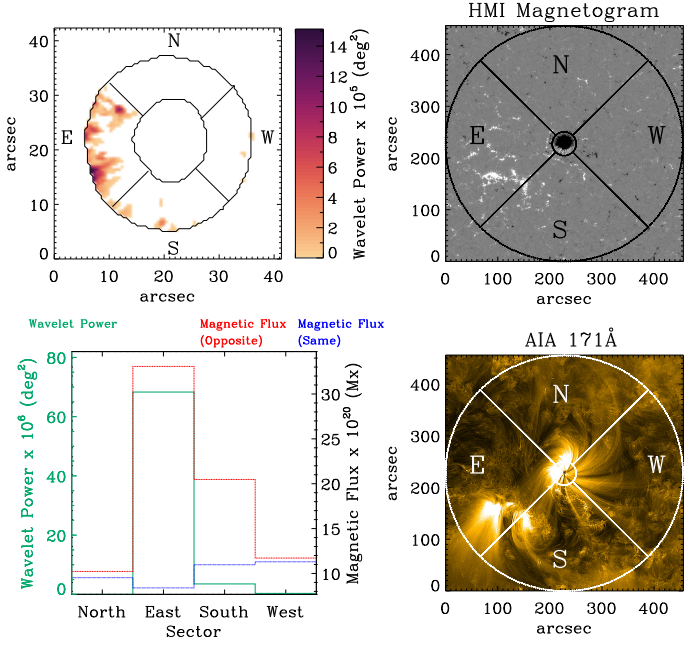}
    \caption{Same as Fig. \ref{Fig:results12218} for the NOAA AR11084.}    
  \label{Fig:results11084}
  \end{center}
\end{figure}

\begin{figure}
  \begin{center}
    \includegraphics[width=0.5\textwidth]{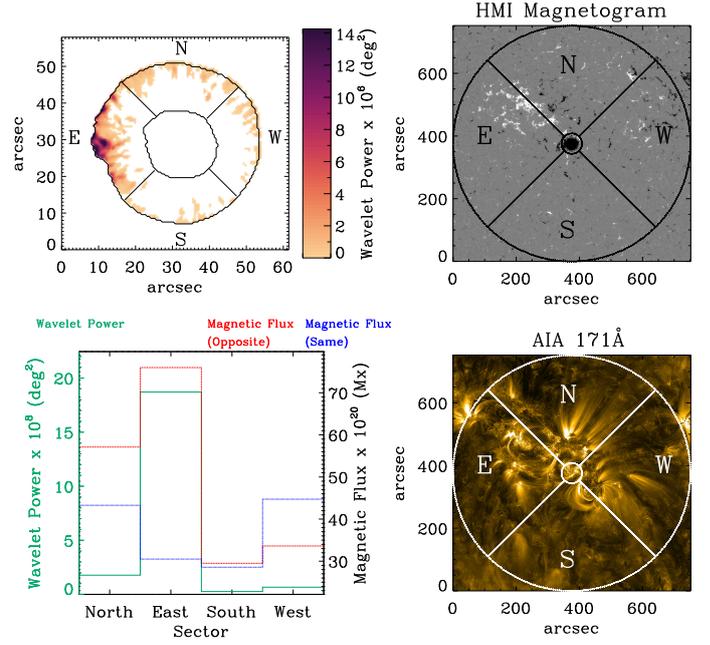}
    \caption{Same as Fig. \ref{Fig:results12218} for the NOAA AR11092.}   
  \label{Fig:results11092}
  \end{center}
\end{figure}

\begin{figure}
  \begin{center}
    \includegraphics[width=0.5\textwidth]{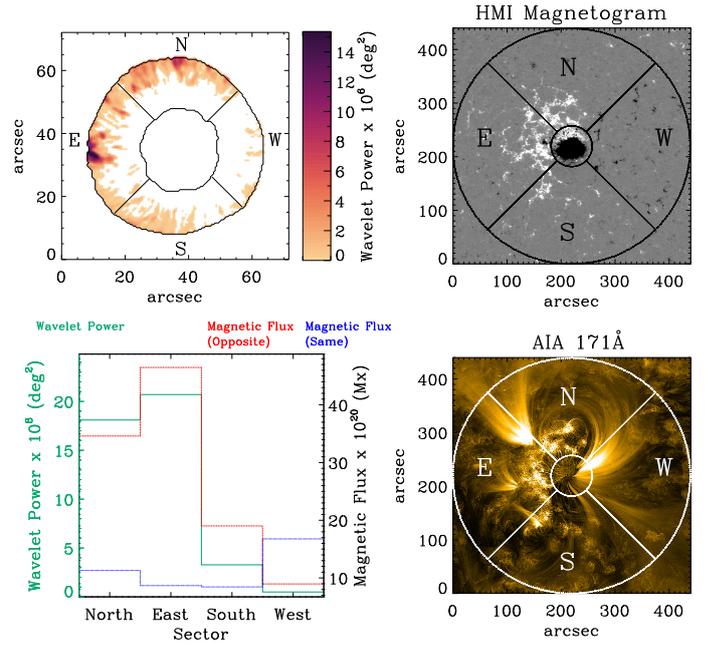}
    \caption{Same as Fig. \ref{Fig:results12218} for the NOAA AR11131.}
    \label{Fig:torsiOsci11131}
  \end{center}
\end{figure}

\begin{figure}
  \begin{center}
    \includegraphics[width=0.5\textwidth]{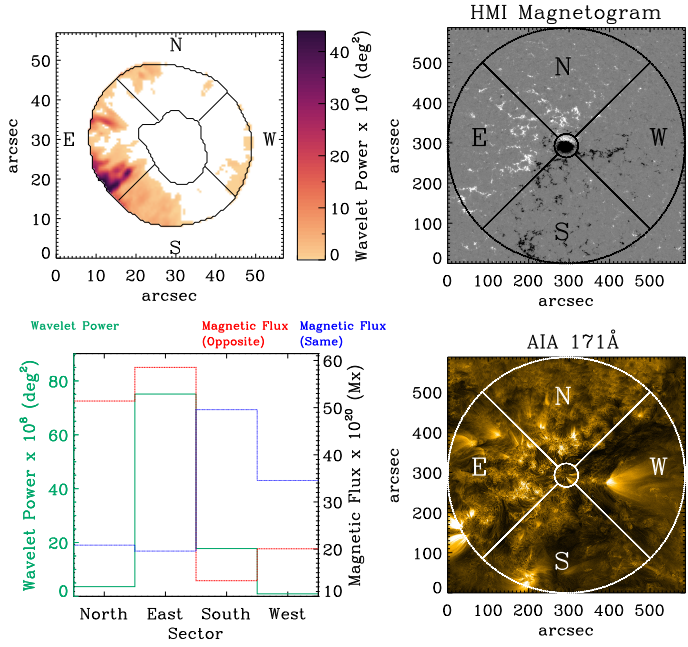}
    \caption{Same as Fig. \ref{Fig:results12218} for the NOAA AR11140.}
    \label{Fig:torsiOsci11140}
  \end{center}
\end{figure}

\begin{figure}
  \begin{center}
    \includegraphics[width=0.5\textwidth]{11216_4panels-eps-converted-to.pdf}
    \caption{Same as Fig. \ref{Fig:results12218} for the NOAA AR11216.}
    \label{Fig:torsiOsci11216}
  \end{center}
\end{figure}

\begin{figure}
  \begin{center}
    \includegraphics[width=0.5\textwidth]{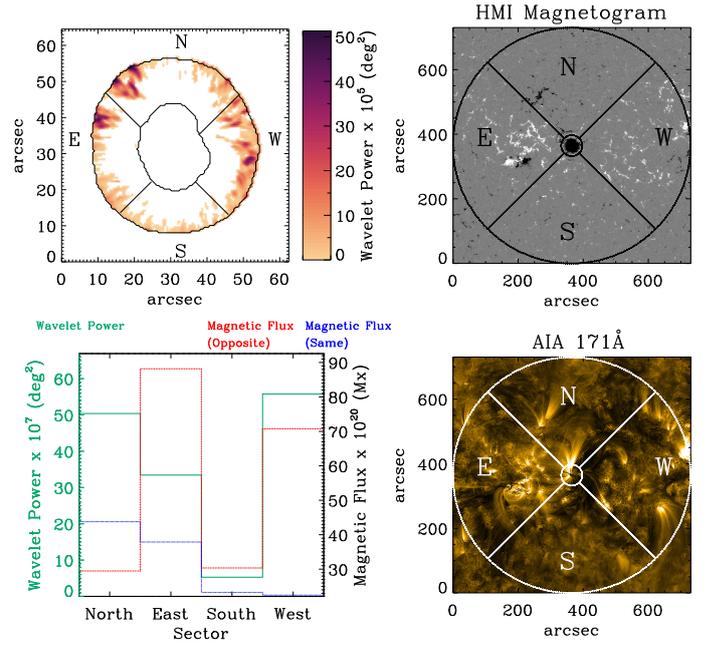}
    \caption{Same as Fig. \ref{Fig:results12218} for the NOAA AR11312.}
    \label{Fig:torsiOsci11312}
  \end{center}
\end{figure}

\begin{figure}
  \begin{center}
    \includegraphics[width=0.5\textwidth]{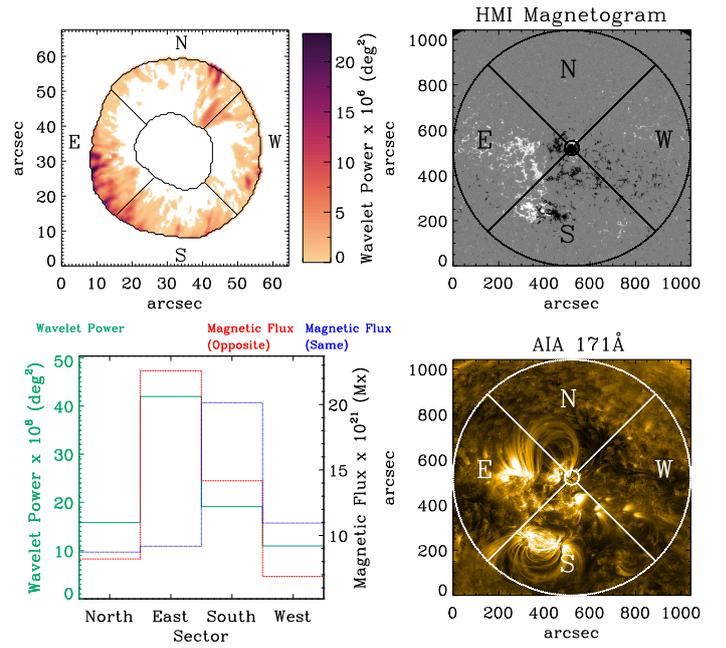}
    \caption{Same as Fig. \ref{Fig:results12218} for the NOAA AR11314.}
    \label{Fig:torsiOsci11314}
  \end{center}
\end{figure}

\begin{figure}
  \begin{center}
    \includegraphics[width=0.5\textwidth]{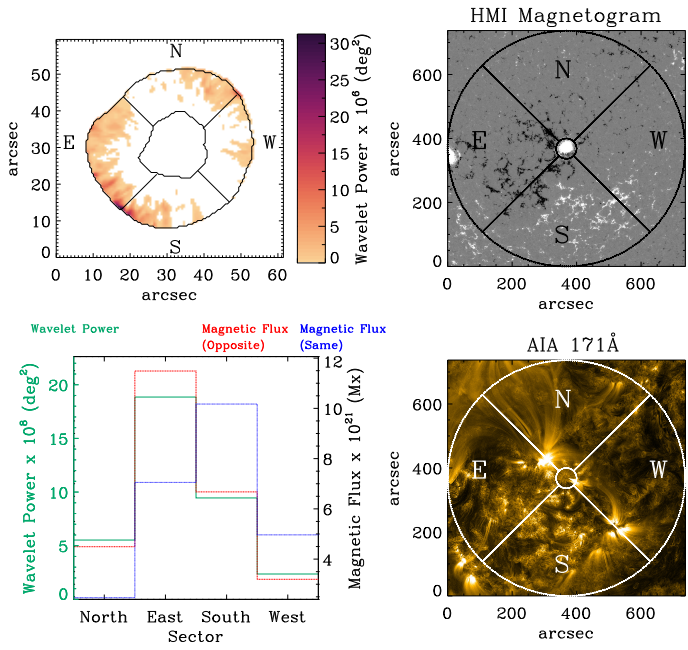}
    \caption{Same as Fig. \ref{Fig:results12218} for the NOAA AR11579.}
    \label{Fig:torsiOsci11579}
  \end{center}
\end{figure}

\begin{figure}
  \begin{center}
    \includegraphics[width=0.5\textwidth]{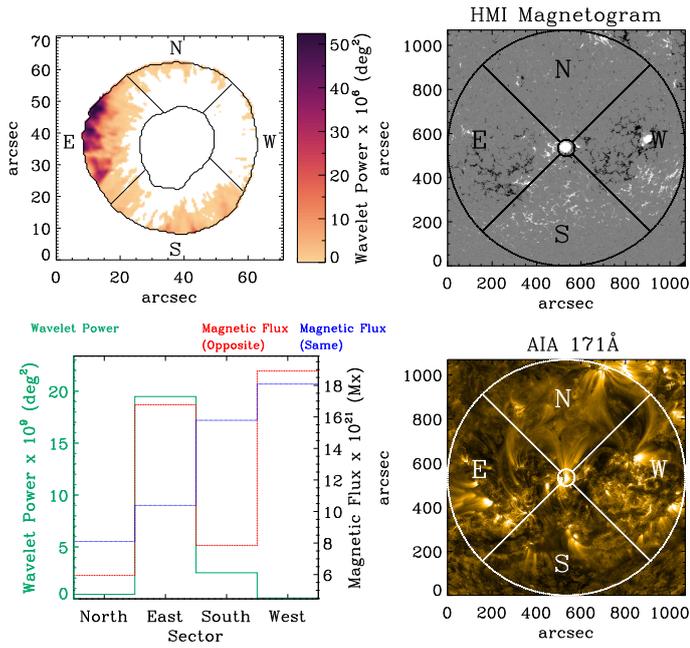}
    \caption{Same as Fig. \ref{Fig:results12218} for the NOAA AR11582.}
    \label{Fig:torsiOsci11582}
  \end{center}
\end{figure}

\begin{figure}
  \begin{center}
    \includegraphics[width=0.5\textwidth]{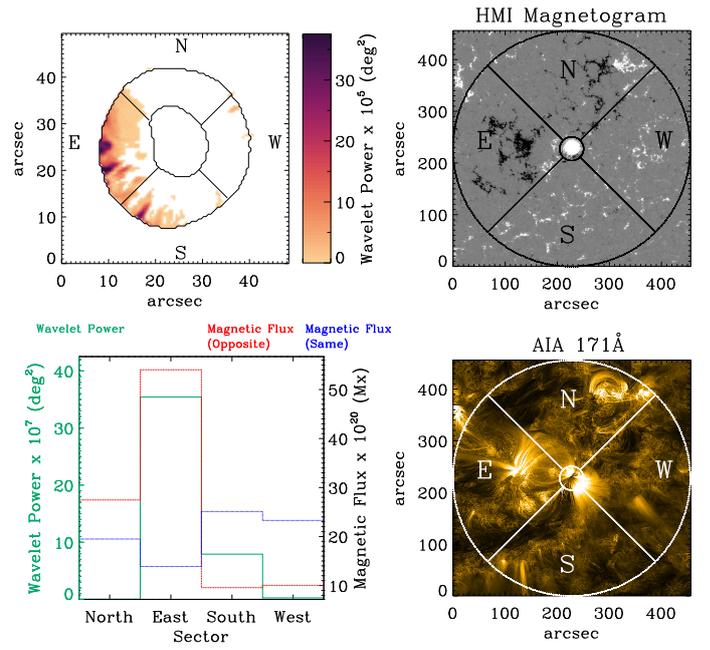}
    \caption{Same as Fig. \ref{Fig:results12218} for the NOAA AR11642.}
    \label{Fig:torsiOsci11642}
  \end{center}
\end{figure}

\begin{figure}
  \begin{center}
    \includegraphics[width=0.5\textwidth]{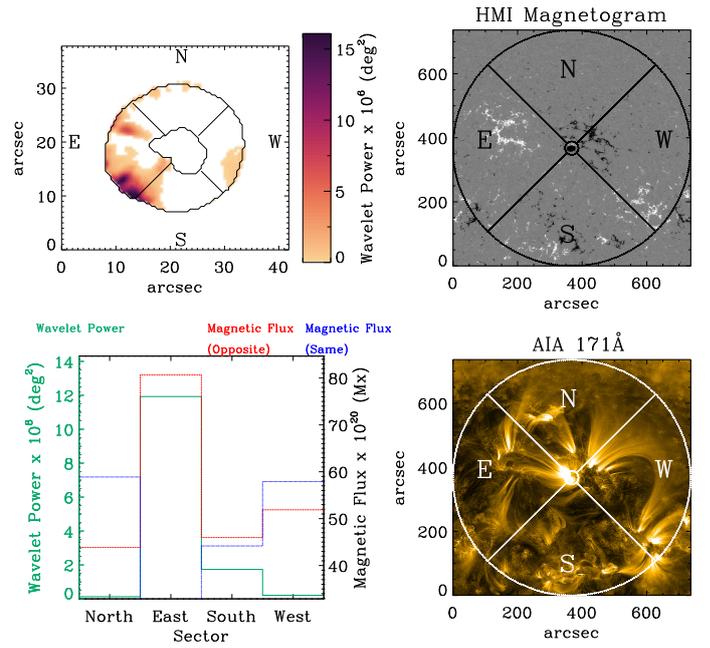}
    \caption{Same as Fig. \ref{Fig:results12218} for the NOAA AR11662.}
    \label{Fig:torsiOsci11662}
  \end{center}
\end{figure}

\begin{figure}
  \begin{center}
    \includegraphics[width=0.5\textwidth]{11665_4panels-eps-converted-to.pdf}
    \caption{Same as Fig. \ref{Fig:results12218} for the NOAA AR11665.}
    \label{Fig:torsiOsci11665}
  \end{center}
\end{figure}

\begin{figure}
  \begin{center}
    \includegraphics[width=0.5\textwidth]{11777_4panels-eps-converted-to.pdf}
    \caption{Same as Fig. \ref{Fig:results12218} for the NOAA AR11777.}
    \label{Fig:torsiOsci11777}
  \end{center}
\end{figure}

\begin{figure}
  \begin{center}
    \includegraphics[width=0.5\textwidth]{11801_4panels-eps-converted-to.pdf}
    \caption{Same as Fig. \ref{Fig:results12218} for the NOAA AR11801.}
    \label{Fig:torsiOsci11801}
  \end{center}
\end{figure}

\begin{figure}
  \begin{center}
    \includegraphics[width=0.5\textwidth]{11846_4panels-eps-converted-to.pdf}
    \caption{Same as Fig. \ref{Fig:results12218} for the NOAA AR11846.}
    \label{Fig:torsiOsci11846}
  \end{center}
\end{figure}

\begin{figure}
  \begin{center}
    \includegraphics[width=0.5\textwidth]{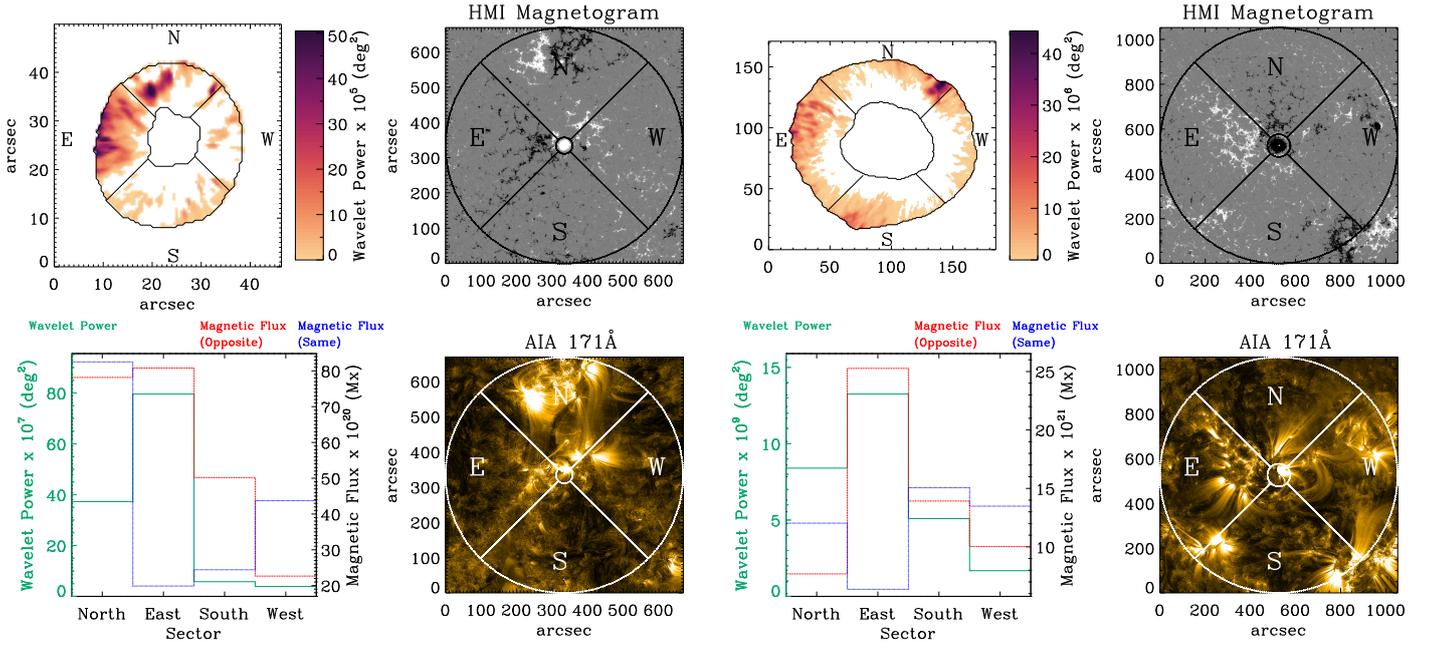}
    \caption{Same as Fig. \ref{Fig:results12218} for the NOAA AR11857.}
    \label{Fig:torsiOsci11857}
  \end{center}
\end{figure}

\begin{figure}
  \begin{center}
    \includegraphics[width=0.5\textwidth]{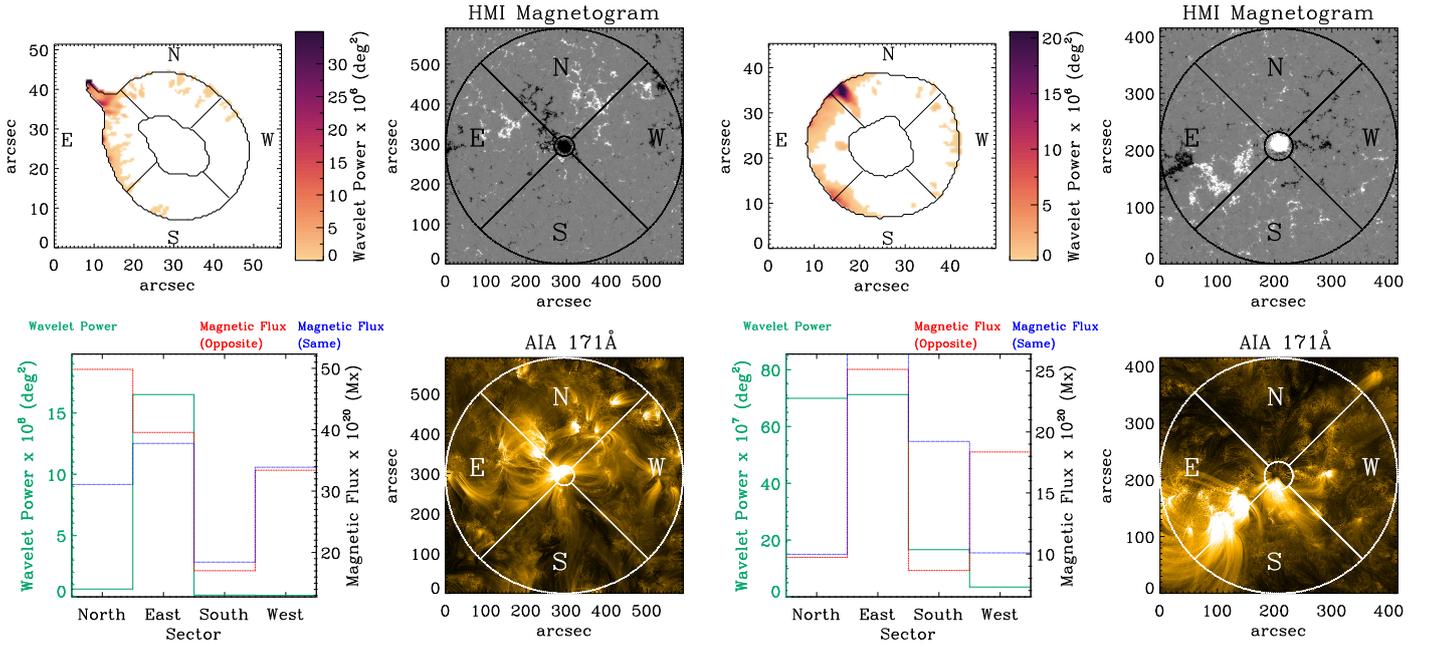}
    \caption{Same as Fig. \ref{Fig:results12218} for the NOAA AR11896.}
    \label{Fig:torsiOsci11896}
  \end{center}
\end{figure}

\begin{figure}
  \begin{center}
    \includegraphics[width=0.5\textwidth]{11899_4panels-eps-converted-to.pdf}
    \caption{Same as Fig. \ref{Fig:results12218} for the NOAA AR11899.}
    \label{Fig:torsiOsci11896}
  \end{center}
\end{figure}

\begin{figure}
  \begin{center}
    \includegraphics[width=0.5\textwidth]{11912_4panels-eps-converted-to.pdf}
    \caption{Same as Fig. \ref{Fig:results12218} for the NOAA AR11912.}
    \label{Fig:torsiOsci11912}
  \end{center}
\end{figure}

\begin{figure}
  \begin{center}
    \includegraphics[width=0.5\textwidth]{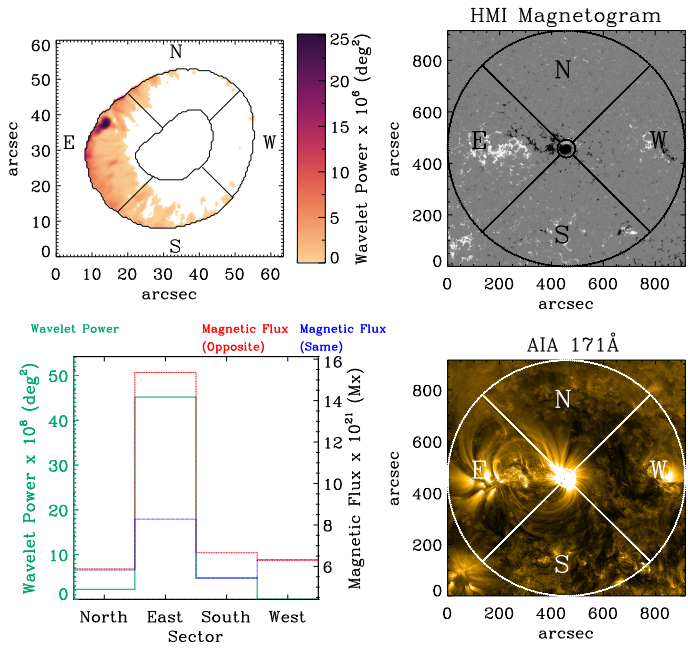}
    \caption{Same as Fig. \ref{Fig:results12218} for the NOAA AR12005.}
    \label{Fig:torsiOsci12005}
  \end{center}
\end{figure}

\begin{figure}
  \begin{center}
    \includegraphics[width=0.5\textwidth]{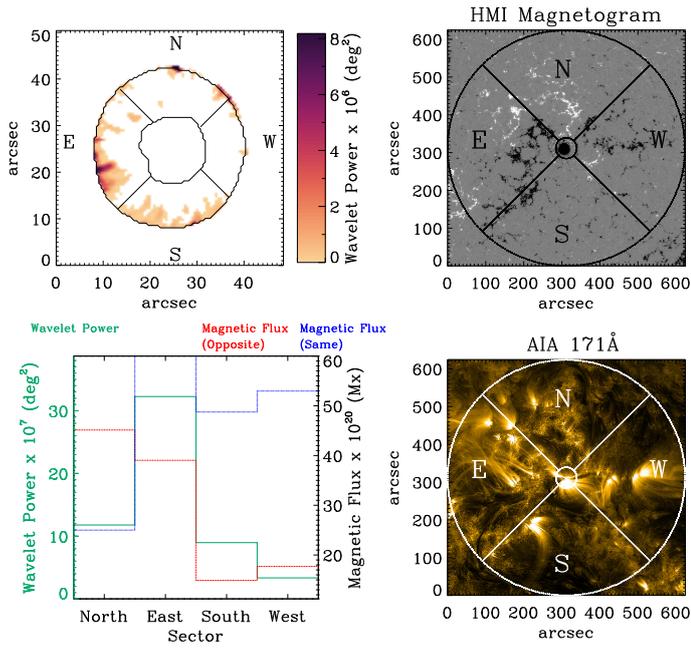}
    \caption{Same as Fig. \ref{Fig:results12218} for the NOAA AR12079.}
    \label{Fig:torsiOsci12079}
  \end{center}
\end{figure}

\begin{figure}
  \begin{center}
    \includegraphics[width=0.5\textwidth]{12186_4panels-eps-converted-to.pdf}
    \caption{Same as Fig. \ref{Fig:results12218} for the NOAA AR12186.}
    \label{Fig:torsiOsci12186}
  \end{center}
\end{figure}

\begin{figure}
  \begin{center}
    \includegraphics[width=0.5\textwidth]{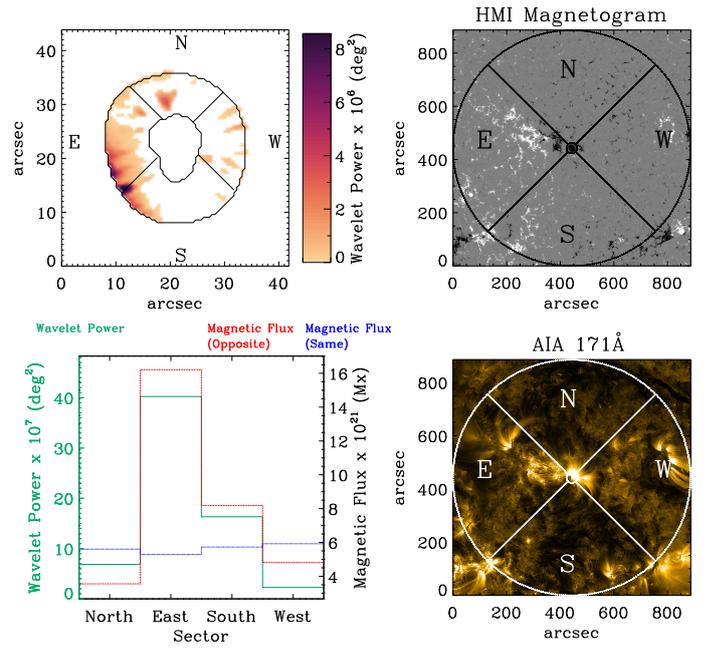}
    \caption{Same as Fig. \ref{Fig:results12218} for the NOAA AR12195.}
    \label{Fig:torsiOsci12195}
  \end{center}
\end{figure}

\begin{figure}
  \begin{center}
    \includegraphics[width=0.5\textwidth]{12218_4panels-eps-converted-to.pdf}
    \caption{Same as Fig. \ref{Fig:results12218}.}
    \label{Fig:torsiOsci12218}
  \end{center}
\end{figure}

\begin{figure}
  \begin{center}
    \includegraphics[width=0.5\textwidth]{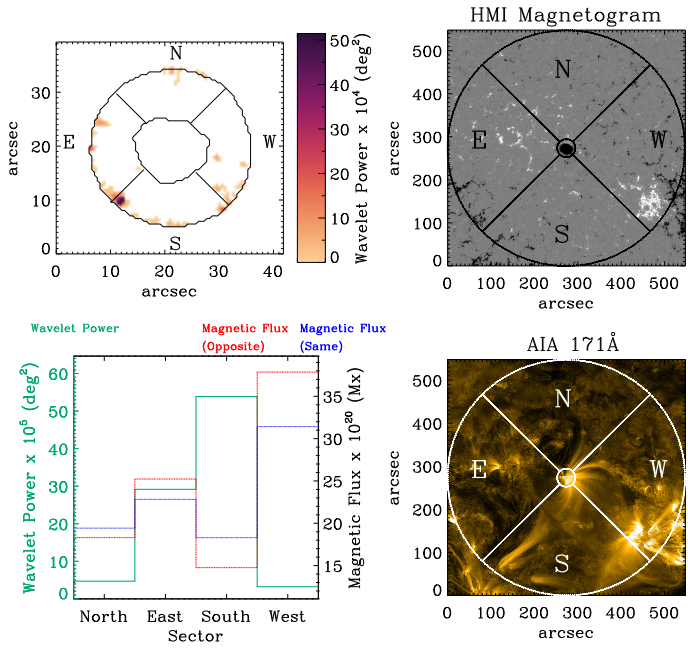}
    \caption{Same as Fig. \ref{Fig:results12218} for the NOAA AR12246 and same as Fig. \ref{Fig:results12246}.}
    \label{Fig:torsiOsci12246}
  \end{center}
\end{figure}

\end{appendix}

\end{document}